\documentstyle[12pt,epsf]{article}
\begin{document}
\large
%title
\begin{titlepage}
\begin{center}
{\large{\bf  { Ground state correlations and
anharmonicity of vibrations}}}\\[1cm]
\end{center}

\begin{center}
A.P. Severyukhin$^1$, V.V. Voronov$^1$, D. Karadjov$^2$\\[2mm]
\end{center}

\noindent
1 - Bogoliubov Laboratory of Theoretical Physics, JINR, Dubna\\
\noindent
2 - Institute for Nuclear Research and Nuclear Energy, Sofia,
Bulgaria\\[5mm]

%abstract
\begin {quotation}
A consistent treatment of the ground state correlations beyond the
random phase approximation
including their influence on the pairing and phonon-phonon coupling
in nuclei is  presented.
A new general system of nonlinear equations for the
quasiparticle phonon model (QPM) is derived.
It is shown that this system contains as a particular case all equations
derived for the QPM early.
New additional Pauli principle corrections resulting in
the anharmonical shifts of energies of the two-phonon configurations are
found.
A correspondence between the generalized QPM equations and the nuclear
field theory is discussed.
\end{quotation}

\vspace{2cm}
\end{titlepage}

\section{Introduction}
Many properties of the nuclear vibrational states can be described
within the Random Phase Approximation (RPA), which enables one to
treat some correlations in the ground state
%\cite{R70,VG71,BM75,Schuck,VG}.
[1-5].
The low-lying nuclear vibrational states investigated with new
generation of detectors \cite{YEH} and the double giant dipole
resonances observed in relativistic heavy ion collision \cite{E94,GSI}
provide an  excellent test for the studies of deviations from the
harmonic picture for multi-phonon excitations in real physical systems.
It is well known that due to the anharmonicity of vibrations there is a
coupling between one-phonon and more complex states \cite{BM75,VG}.
Usually such a coupling was considered for the RPA phonons only \cite{VG}.

From another point of view the RPA violates the
Pauli principle and many attempts have been done to improve it
[9-29].
Extended RPA equations include additional corrections for the ground
state correlations (GSC) \cite{KVC93,KVC94,SVK99,VKCS00}.
In the papers \cite{KVCGS98,SVK99,VKCS00} phonons of the extended RPA
containing corrections for the GSC have been used  as a basis on which
the quasiparticle-phonon model (QPM) equations \cite{VG,AI,VV} are
generalized.
The influence of the GSC on properties of nuclear vibrational states
constructed by  one- and two-phonon configurations was
studied in \cite{KVCGS98}.
Besides the GSC,  the Pauli principle corrections arising in the
two-phonon terms due to the fermion structure of
the phonon operators were taken into account and
the particle-particle channel was included too  \cite{SVK99,VKCS00}.

However, these extended QPM equations have been  derived for the
case of the so called quasidiagonal approximation for the Pauli
principle corrections.
In the present studies  we don't use this approximation.
The second improvement seems more important.
Namely, it is found that there are  additional Pauli principle
corrections resulting in the anharmonicity shifts of the two-phonon
configuration energies.
A correspondence between the anharmonicity shifts
of the QPM and  Nuclear Field Theory diagrams is
discussed.
\section{Hamiltonian of the model}
We employ the QPM hamiltonian \cite{VG,AI,VV}
including an average nuclear field described
as the Woods-Saxon potential, pairing interactions, the isoscalar and
isovector particle--hole (p--h)  and
particle--particle (p--p)
residual forces in a separable form
with the
Bohr--Mottelson radial dependence \cite{BM75}:
\begin{eqnarray}
H &=&\sum\limits_\tau \left( \left. \sum\limits_{jm}\right. ^\tau
(E_j-\lambda _\tau )a_{jm}^{\dagger }a_{jm}-\frac 14G_\tau
^{(0)}:P_0^{\dagger }\,(\tau )P_0\,(\tau ):-\right.  \nonumber\\
&&\ \ \left. \frac 12\sum_a\sum_{\sigma =\pm 1}\sum\limits_{\lambda \mu
}\left[ \left( \kappa _0^{(\lambda ,a)}+\sigma \kappa _1^{(\lambda
,a)}\right) :M_{\lambda \mu }^{\left( a\right) +}(\tau )M_{\lambda \mu
}^{\left( a\right) }(\sigma \tau ):\right] \right)\label{1},
\end{eqnarray}
We sum over the proton($p$) and neutron($n$) indexes and the notation $%
\{\tau =(n,p)\}$ is used and a change $\tau \leftrightarrow -\tau $ means a
change $p\leftrightarrow n$; $a$ is the channel index $a=\{ph,pp\}$.
The single-particle states are specified by the
quantum numbers $(jm)$; $E_j$ are the single-particle energies; $\lambda
_\tau $ is the chemical potential; $G_\tau ^{(0)}$ and $\kappa ^{(\lambda )}$
are the strengths in the p--p and in the p--h channel, respectively. The
monopole pair creation and the multipole operators entering the normal
products in (\ref{1}) are defined as follows:
$$
P_0^{+}\,(\tau )=\,\left. \sum_{jm}\right. ^\tau
(-1)^{j-m}a_{jm}^{+}a_{j-m}^{+},
$$
\[
M_{\lambda \mu }^{\left( ph\right) +}\left( \tau \right) \,=\,\frac 1{\sqrt{
2\lambda +1}}{}\left. \sum_{jj^{^{\prime }}mm^{^{\prime }}}\right. ^\tau
(-1)^{j+m}\langle jmj^{^{\prime }}-m^{^{\prime }}\mid \lambda \mu \rangle
f_{j^{\prime }j}^{(\lambda )}a_{jm}^{+}a_{j^{^{\prime }}m^{^{\prime }}},
\]
\[
M_{\lambda \mu }^{\left( pp\right) +}\left( \tau \right) \,=\,\frac{\left(
-1\right) ^{\lambda -\mu }}{\sqrt{2\lambda +1}}\left. \sum_{jj^{^{\prime
}}mm^{^{\prime }}}\right. ^\tau \langle jmj^{^{\prime }}m^{^{\prime }}\mid
\lambda \mu \rangle f_{jj^{\prime }}^{(\lambda )}a_{jm}^{+}a_{j^{^{\prime
}}m^{^{\prime }}}^{+},
\]
where $f_{jj^{^{\prime }}}^{(\lambda )}$ are the single particle radial
matrix elements of residual forces.
\section{Extended RPA}
In what follows we work in the quasiparticle  representation, defined by
the canonical Bogoliubov transformation:
{\large
\begin{equation}  \label{BG}
a_{jm}^{+}\,=\,u_j\alpha _{jm}^{+}\,+\,(-1)^{j-m}v_j\alpha _{j-m}.
\end{equation}
}
The Hamiltonian (\ref{1}) can be represented in terms of bifermion
quasiparticle operators (and their conjugate ones) \cite{VG,AI,VV}:
\[
B(jj^{^{\prime }};\lambda \mu )\,=\,\sum_{mm^{^{\prime }}}(-1)^{j^{^{\prime
}}+m{^{\prime }}}\langle jmj^{^{\prime }}m^{^{\prime }}\mid \lambda \mu
\rangle \alpha _{jm}^{+}\alpha _{j^{^{\prime }}-m^{^{\prime }}},
\]
\[
A^{+}(jj^{^{\prime }};\lambda \mu )\,=\,\sum_{mm^{^{\prime }}}\langle
jmj^{^{\prime }}m^{^{\prime }}\mid \lambda \mu \rangle \alpha
_{jm}^{+}\alpha _{j^{^{\prime }}m^{^{\prime }}}^{+}.
\]
We introduce the phonon creation operators
{\large
\begin{equation}  \label{BGB}
Q_{\lambda \mu i}^{+}\,=\,\frac 12\sum_{jj^{^{\prime }}}\left( \psi
_{jj^{^{\prime }}}^{\lambda i}\,A^{+}(jj^{^{\prime }};\lambda \mu
)-(-1)^{\lambda -\mu }\varphi _{jj^{^{\prime }}}^{\lambda i}\,A(jj^{^{\prime
}};\lambda -\mu )\right).
\end{equation}
}
where the index $\lambda $ denotes multipolarity and $\mu $ is
its z-projection in the laboratory system.
One assumes that the ground state  is the ERPA phonon vacuum
$\mid 0\rangle $, i.e. $Q_{\lambda \mu i}\mid 0\rangle\,=0$.
We define the excited states for this approximation by
$Q_{\lambda\mu i}^{+}\mid0\rangle$.
The following relation can be
proved using the exact commutators of the fermion operators:
{\large
\[
\langle 0\mid \,[Q_{\lambda \mu ,i},Q_{\lambda ^{^{\prime }}\mu ^{^{\prime
}},i^{^{\prime }}}^{+}]\,\mid 0\rangle \,=\\
\frac 12\sum_{jj^{^{\prime }}}\,(1\,-q_{jj^{^{\prime }}})\left( \psi
_{jj^{^{\prime }}}^{\lambda i}\,\psi _{jj^{^{\prime }}}^{\lambda i^{^{\prime
}}}\,-\,\varphi _{jj^{^{\prime }}}^{\lambda i}\varphi _{jj^{^{\prime
}}}^{\lambda i^{^{\prime }}}\right),
\]
}
where $q_{jj^{^{\prime}}}=q_j+q_{j^{^{\prime }}}$ and $q_j$
is the quasiparticle distribution in
the ground state:
$
q_j= \sum_m\langle 0\mid \alpha _{jm}^{+}\alpha _{jm}\mid 0\rangle
\left( 2j+1\right) ^{-1}.
$
The quasiparticle energies ($\varepsilon _j$), the energies of
the ERPA excited states ($\omega _{_{\lambda i}}$),
the chemical potentials ($\lambda_\tau $), the coefficients
($u$,$v$,$\psi$,$\varphi$) of the  Bogoliubov transformations
(\ref{BG}), (\ref{BGB})  and  the
quasiparticle distributions in the ground state ($q_j$) are determined
from  the following non--linear system of equations obtained by using
the general equation of motion approach \cite{R70,R2}:
{\large
\begin{equation}  \label{S1}
\langle 0|\left\{ \delta \alpha _{jm},\left[ H,\alpha _{jm}^{+}\right]
\right\} \mid 0\rangle =\varepsilon _j\langle 0|\left\{ \delta \alpha
_{jm},\alpha _{jm}^{+}\right\} \mid 0\rangle,
\end{equation}
}
{\large
\begin{equation}  \label{S4}
\langle 0|\left[ \delta Q_{\lambda \mu i},\left[ H,Q_{\lambda \mu
i}^{+}\right] \right] \mid 0\rangle =\omega _{_{\lambda i}}\langle 0|\left[
\delta Q_{\lambda \mu i},Q_{\lambda \mu i}^{+}\right] \mid 0\rangle,
\end{equation}
}
with the constrained conditions:
{\large
\begin{equation}  \label{S2}
\left\{ \alpha _{jm},\alpha _{j^{\prime }m^{\prime }}^{+}\right\} =\delta
_{jm,j^{\prime }m^{\prime }},
\end{equation}
}
{\large
\begin{equation}  \label{S3}
\langle 0|\left. \sum_{jm}\right. ^\tau a_{jm}^{+}a_{jm}|0\rangle =N^{\left(
\tau \right) } ,
\end{equation}
}
{\large
\begin{equation}  \label{S5}
\langle 0\mid \,[Q_{\lambda \mu i},Q_{\lambda \mu i^{^{\prime }}}^{+}]\,\mid
0\rangle \,=\delta _{ii^{^{\prime }}},
\end{equation}
}
{\large
\begin{equation}
q_j=\,\frac 12\sum_{\lambda i,j^{^{\prime }}}\left( \frac{2\lambda \,+\,1}{
2j\,+\,1}\right) \,\,(1\,-q_{jj^{^{\prime }}})\,\left( \varphi
_{jj^{^{\prime }}}^{\lambda i}\right) ^2 ,\label{S6}
\end{equation}
}
where $N^{\left(\tau \right) }$  is
the  number of particles.
The variations $\delta \alpha _{jm}$ and $\delta Q_{\lambda \mu i}$
have the following  form:
{\large
\[
\delta \alpha _{jm}\,=\,\delta u_ja_{jm}\,-\,(-1)^{j-m}\delta
v_ja_{j-m}^{+},
\]
}
{\large
\[
\delta Q_{\lambda \mu i}\,=\,\frac 12\sum_{jj^{^{\prime }}}\left( \delta
\psi _{jj^{^{\prime }}}^{\lambda i}\,A(jj^{^{\prime }};\lambda \mu
)-(-1)^{\lambda -\mu }\delta \varphi _{jj^{^{\prime }}}^{\lambda
i}\,A^{+}(jj^{^{\prime }};\lambda -\mu )\right).
\]
}
The system of equations (\ref{S1})-(\ref{S6}) can be derived using
various approaches, e.g. \cite{JR}.
For $q_j=0$ the system of equations (\ref{S1})-(\ref{S6})
reduces to the usual BCS and RPA equations with
the p-h and p-p channels \cite{VG,AI}.

The equation (\ref{S4}) have been obtained under the assumptions:
{\large
\begin{equation}
\langle 0|\left\{ \alpha _{j_1^{\prime }m_1^{\prime }},\left[ H,\alpha
_{jm}\right] \right\} \alpha _{j^{\prime }m^{\prime }}\alpha _{j_1m_1}\mid
0\rangle \approx 0,  \label{A1}
\end{equation}
}
{\large
\begin{equation}
\langle 0|\left[ H,\alpha _{jm}^{+}\right] \alpha _{j_1m_1}\mid 0\rangle
\approx \varepsilon _jq_j\delta _{jm,j_1m_1},  \label{A2}
\end{equation}
}
{\large
\begin{equation}
\langle 0|\left\{ \left[ H,\alpha _{jm}^{+}\right] ,\alpha _{j_1m_1}\right\}
\alpha _{j^{\prime }m^{\prime }}^{+}\alpha _{j_1^{\prime }m_1^{\prime }}\mid
0\rangle \approx \varepsilon _jq_{j^{\prime }}\delta _{jm,j_1m_1}\delta
_{j^{\prime }m^{\prime },j_1^{\prime }m_1^{\prime }}.  \label{A3}
\end{equation}
}
If we put $q_j=0$ then these approximations correspond to the procedure
of the linearization of equations.
One can derive the equation  (\ref{S6}) using various methods
\cite{R1,LW90}.
Using the completeness and orthogonality conditions for the
phonon operators one can express bifermion operators
$A^{+}(jj^{^{\prime }};\lambda \mu )$ and  $A(jj^{^{\prime }};\lambda
\mu)$  as
{\large
\begin{equation}
\label{REV}
A^{+}(jj{^{\prime }};\lambda \mu )=(1\,-\,q_{jj^{^{\prime }}})\sum_i\left(
\psi _{jj^{^{\prime }}}^{\lambda i}Q_{\lambda \mu i}^{+}\,+\,(-)^{\lambda
-\mu }\,\varphi _{jj^{^{\prime }}}^{\lambda i}Q_{\lambda -\mu i}\right)
\end{equation}
}
It is necessary to point out that
the solutions of the system of equations
(\ref{S1})-(\ref{S6}) obey the following equalities:
{\large
\begin{equation}
\label{ERPA}
\left. \langle 0|\left[ Q_{\lambda \mu i}\,,\left[ H\,,Q_{\lambda \mu
i^{\prime }}^{+}\right] \right] |0\rangle \right. _{ERPA}=\,\omega
_{\lambda i}\delta _{ii^{\prime }},
\end{equation}
}
{\large
\[
\label{QQ}
\left. \langle 0|HQ_{\lambda _1\mu _1i_1}^{+}Q_{\lambda _2\mu
_2i_2}^{+}|0\rangle \right. _{ERPA}=0.
\]
}
The proof of these statements is analogous to that in the usual RPA case
\cite{VG71}.

As it was shown in Refs.\cite{KVC93,KVCGS98,VKCS00}  the ERPA
calculations give a better
agreement with experimental data for the characteristics of
the low-lying states than the RPA ones.

\section{Generalized QPM equations}
The GSC affect not only the RPA, but they also should change  \\the
quasiparticle-phonon coupling (see Refs.\cite{KVCGS98,SVK99,VKCS00}).
To take into account such effects we follow the basic ideas of the
QPM. Hereafter we generalize the extended QPM equations
\cite{KVCGS98,SVK99,VKCS00}.
As it was shown in \cite{KVC96} the pairing vibrations give a negligible
contribution to $q_j$. On the other hand the two-phonon configurations
including the pairing vibration phonons have an energy essentially higher
than the configurations constructed from usual vibration phonons. That is
why we do not take into account the coupling with the pairing vibrations
($\lambda=0$) in what follows.

The initial Hamiltonian (\ref{1}) can be rewritten in terms of
quasiparticle and phonon operators in the following form:
{\large
\begin{equation}
H\,=\,h_0\,+\,h_0^{_{pp}}\,+\,h_{QQ}\,+\,h_{QB},
\label{H1}
\end{equation}
}
{\large
\[
h_0\,+\,h_0^{_{pp}}\,=\,\sum_{jm}\,\varepsilon _j\,\alpha _{jm}^{+}\,\alpha
_{jm},
\]
}
{\large
\[
h_{QQ}=h_1+h_2+h_3  ,
\]
}
{\large
\[
h_1=-\frac 14\sum_{\lambda \mu ii^{^{\prime }}\tau }\frac{\overline{X}%
^{\lambda ii^{\prime }}\left( \tau \right) +\overline{X}^{\lambda i^{\prime
}i}\left( \tau \right) }{\sqrt{{\cal Y}_\tau ^{\lambda i}{\cal Y}_\tau
^{\lambda i^{^{\prime }}}}}Q_{\lambda \mu i}^{+}Q_{\lambda \mu i^{^{\prime
}}},
\]
}
{\large
\[
h_2=-\frac 18\sum_{\lambda \mu ii^{^{\prime }}\tau }\frac{{\cal Z}^{\lambda
ii^{\prime }}\left( \tau \right) }{\sqrt{{\cal Y}_\tau ^{\lambda i}{\cal Y}%
_\tau ^{\lambda i^{^{\prime }}}}}\left( -1\right) ^{\lambda -\mu }\left(
Q_{\lambda \mu i}^{+}Q_{\lambda -\mu i^{^{\prime }}}^{+}+Q_{\lambda -\mu
i^{^{\prime }}}Q_{\lambda \mu i}\right),
\]
}
{\large
\[
h_3=-\frac 18\sum_{\lambda \mu ii^{^{\prime }}\tau }\frac{\overline{X}%
_1^{\lambda ii^{\prime }}\left( \tau \right) +\overline{X}_1^{\lambda
i^{\prime }i}\left( \tau \right) }{\sqrt{{\cal Y}_\tau ^{\lambda i}{\cal Y}%
_\tau ^{\lambda i^{^{\prime }}}}}\left[ Q_{\lambda \mu i},Q_{\lambda \mu
i^{^{\prime }}}^{+}\right],
\]
}
{\large
\begin{eqnarray*}
h_{QB}=-\frac 1{2\sqrt{2}}\sum_{\lambda \mu i\tau }\left. \sum_{jj^{^{\prime
}}}\right. ^\tau \frac{f_{jj^{^{\prime }}}^{(\lambda )}}{\sqrt{{\cal Y}_\tau
^{\lambda i}}}\left( (-)^{\lambda -\mu }Q_{\lambda \mu i}^{+}L_{jj^{^{\prime
}}}^{\lambda i\left( +\right) }\left( \tau \right) +\right.  \\
\left. Q_{\lambda -\mu i}L_{jj^{^{\prime }}}^{\lambda i\left( -\right)
}\left( \tau \right) \right) B(jj^{^{\prime }};\lambda -\mu )+h.c. ,
\end{eqnarray*}
}
The coefficients of the hamiltonian (\ref{H1}) are given in Appendix A.
The term $h_{QB}$ is responsible for the mixing of the configurations.
While constructing  the hamiltonian
(\ref{H1})  we have neglected the terms
$
\sim :B(jj^{^{\prime }};\lambda \mu )\,B(j_1j_1^{^{\prime }};\lambda \mu ):
$,
which do not lead to coherent effects and the energy
corrections due to these terms are small in spherical nuclei
\cite{VG,KVKP90}.

The commutation relations
$\left[ \left[ Q_1,Q_2^{+}\right] ,Q_3^{+}\right]$,
$\left[ \left[ Q_1^{+},Q_2^{+}\right] ,Q_3^{+}\right]$ and \\
$\left[ Q_3,\left[ Q_1^{+},Q_2^{+}\right] \right]$ are calculated by
using  the transformation (\ref{REV}).
{\large
\begin{equation}
\left[ \left[ Q_1,Q_2^{+}\right] ,Q_3^{+}\right]
=\sum_4K_1(41|23)Q_4+K(41|23)Q_4^{+}\label{k3.0}
\end{equation}
}
{\large
\begin{equation}
\left[ \left[ Q_1^{+},Q_2^{+}\right] ,Q_3^{+}\right]
=\sum_4K_3(41|23)Q_4+K_2(41|23)Q_4^{+}\label{k3.1}
\end{equation}
}
{\large
\begin{equation}
\left[ Q_3,\left[ Q_1^{+},Q_2^{+}\right] \right]
=\sum_4K_5(41|23)Q_4+K_4(41|23)Q_4^{+}\label{k3.2}
\end{equation}
}
These coefficients are of fourth-order in phonon amplitudes.
In what  follows one needs the coefficients $K$, $K_1$, $K_2$,
$K_4$ only. They are given in Appendix B.
We obtain the anharmonic corrections taking into
account these commutation relations.

Hereinafter the ground state is approximated by the ERPA phonon vacuum.
Using the ERPA phonons as a basis the wave functions
of the excited states  of even--even nuclei can be written as:
{\large
\begin{equation}
\label{WF}
\Psi _\nu (\lambda \mu )=\Omega _{\lambda \mu \nu }^{+}|0\rangle,
\end{equation}
}
where
{\large
\[
\Omega _{\lambda \mu \nu }^{+}=\sum_iR_i(\lambda \nu )Q_{\lambda \mu
i}^{+}+\sum_{\lambda _1i_1\lambda _2i_2}P_{\lambda _2i_2}^{\lambda
_1i_1}(\lambda \nu )\left( Q_{\lambda _1\mu _1i_1}^{+}Q_{\lambda _2\mu
_2i_2}^{+}\right) _{\lambda \mu }\label{f2},
\]
}
{\large
\[
\left( Q_{\lambda _1\mu _1i_1}^{+}Q_{\lambda _2\mu _2i_2}^{+}\right)
_{\lambda \mu }\equiv \sum_{\mu _1\mu _2}C_{\lambda _1\mu _1\lambda _2\mu
_2}^{\lambda \mu }Q_{\lambda _1\mu _1i_1}^{+}Q_{\lambda _2\mu_2i_2}^{+}.
\]
}
The wave functions (\ref{WF}) have the normalization condition
{\large
\begin{eqnarray}
\label{norma}
&&\sum\limits_iR_i^2(J\nu )+\sum_{\lambda _1i_1\lambda _2i_2}\sum_{\lambda
_{1^{\prime }}i_{1^{\prime }}\lambda _{2^{\prime }}i_{2^{\prime
}}}P_{\lambda _2i_2}^{\lambda _1i_1}(J\nu )P_{\lambda _2^{\prime
}i_2^{\prime }}^{\lambda _1^{\prime }i_1^{\prime }}(J\nu )\times  \\ \nonumber
&&\left( \delta _{\lambda _2^{\prime }i_2^{\prime },\lambda _2i_2}\delta
_{\lambda _1^{\prime }i_1^{\prime },\lambda _1i_1}+\delta _{\lambda
_2^{\prime }i_2^{\prime },\lambda _1i_1}\delta _{\lambda _1^{\prime
}i_1^{\prime },\lambda _2i_2}\left( -1\right) ^{\lambda _1+\lambda
_2+J}\right. +
\end{eqnarray}
\[
\left. K^J(\lambda _2^{\prime }i_2^{\prime },\lambda _1^{\prime }i_1^{\prime
}\mid \lambda _1i_1,\lambda _2i_2)\right)=1.
\]
We use the equation of motion method to diagonalize the
hamiltonian (\ref{H1})
{\large
\[
\langle 0|\left[ \delta \Omega _{\lambda \mu \nu },\left[ H,\Omega _{\lambda
\mu \nu }^{+}\right] \right] \mid 0\rangle =E_\nu \langle 0|\left[
\delta \Omega _{\lambda \mu \nu },\Omega _{\lambda \mu \nu }^{+}\right]
|0\rangle,
\]
}
where the variation $\delta \Omega _{\lambda \mu \nu }$ has the
following form
{\large
\[
\delta \Omega _{\lambda \mu \nu }=\sum_i\delta R_i(\lambda \nu )Q_{\lambda
\mu i}+\sum_{\lambda _1i_1\lambda _2i_2}\delta P_{\lambda _2i_2}^{\lambda
_1i_1}(\lambda \nu )\left( Q_{\lambda _2\mu _2i_2}Q_{\lambda _1\mu
_1i_1}\right) _{\lambda \mu }\left( -1\right) ^{\lambda _1+\lambda
_2+\lambda }.
\]
}
The variational principle yields a set of linear equations for the
unknown  wave function coefficients $R_i(J\nu)$ and $P_{\lambda_2i_2}^{\lambda_1i_1}(J\nu )$
{\large
\begin{equation}
\label{EQ}
\left(
\begin{tabular}{ll}
${\cal W}_1$ & ${\cal U}$ \\
${\cal U}^T$ & ${\cal W}_2$%
\end{tabular}
\right) \left(
\begin{tabular}{l}
${\cal R}\left( \nu \right) $ \\
${\cal P}\left( \nu \right) $%
\end{tabular}
\right) =E_\nu \left(
\begin{tabular}{ll}
$I$ & $0$ \\
$0$ & $\overline{{\cal K}}$%
\end{tabular}
\right) \left(
\begin{tabular}{l}
${\cal R}\left( \nu \right) $ \\
${\cal P}\left( \nu \right) $%
\end{tabular}
\right)
\end{equation}
}
with the additional condition (\ref{norma}).
The  matrix ${\cal U}^T$ is a transpose of the matrix $\cal{U}$.
The number of linear equations (\ref{EQ}) equals  to the number of one-
and two-phonon configurations included in the wave function (\ref{WF}).
The notations introduced above are
{\large
\begin{equation}
\left. {\cal W}_1\right. _i^{i^{\prime }}\left( \lambda \right)=\langle
0|\left[ Q_{\lambda \mu i^{\prime }}\,,\left[ H\,,Q_{\lambda \mu
i}^{+}\right] \right] |0\rangle \label{e2},
\end{equation}
}

{\large
\begin{eqnarray}
\left. {\cal W}_2\right. _{\lambda _1i_1,\lambda
_2i_2}^{\lambda _1^{\prime }i_1^{\prime },\lambda _2^{\prime}i_2^{\prime }}
\left( \lambda \right)&=&\left( -1\right) ^{\lambda _1^{\prime }+\lambda _2^{\prime }+\lambda
}\times   \label{e3} \\\nonumber
&&\ \langle 0|\left[ \left( Q_{\lambda _2^{\prime }\mu _2^{\prime
}i_2^{\prime }}Q_{\lambda _1^{\prime }\mu _1^{\prime }i_1^{\prime }}\right)
_{\lambda \mu },\left[ H,\left( Q_{\lambda _1\mu _1i_1}^{+}Q_{\lambda _2\mu
_2i_2}^{+}\right) _{\lambda \mu }\right] \right] |0\rangle,
\end{eqnarray}
}
{\large
\begin{equation}
{\cal U}_{\lambda _1i_1,\lambda
_2i_2}^{\lambda i}\left( \lambda \right)=\langle 0|\left[ Q_{\lambda \mu i},\left[ H,\left(
Q_{\lambda _1\mu _1i_1}^{+}Q_{\lambda _2\mu _2i_2}^{+}\right) _{\lambda \mu
}\right] \right] |0\rangle \label{e4},
\end{equation}
}
{\large
\[
I_i^{i^{\prime }}=\delta_{i^{\prime },i}\label{e5},
\]
}
{\large
\begin{eqnarray*}
\left. \overline{{\cal K}}\right. _{\lambda
_1i_1,\lambda _2i_2}^{\lambda _1^{\prime }i_1^{\prime},\lambda_2^{\prime
}i_2^{\prime }}\left( \lambda \right) &=&\left( \delta _{\lambda
_2^{\prime }i_2^{\prime },\lambda
_2i_2}\delta _{\lambda _1^{\prime }i_1^{\prime },\lambda _1i_1}+\delta
_{\lambda _2^{\prime }i_2^{\prime },\lambda _1i_1}\delta _{\lambda
_1^{\prime }i_1^{\prime },\lambda _2i_2}\left( -1\right) ^{\lambda
_1+\lambda _2+\lambda}\right. +  \label{e6} \\
&&\ \left. K^{\lambda}(\lambda _2^{\prime }i_2^{\prime },\lambda
_1^{\prime }i_1^{\prime }\mid \lambda _1i_1,\lambda _2i_2)\right).
\end{eqnarray*}
}
Now we calculate the matrix elements  (\ref{e2}), (\ref{e3}) and as a
result we get
{\large
\begin{equation}
\label{w1}
\left. {\cal W}_1\right. _{i_1}^{i_2^{\prime
}}\left( \lambda _1\right)=\delta _{i_2^{\prime },i_1}\omega _{\lambda _1i_1}-\frac 18\sum_{\lambda
J^{\prime }ii^{^{\prime }}\tau }\left( \frac{2J^{\prime }+1}{2\lambda _1+1}%
\right) \frac 1{\sqrt{{\cal Y}_\tau ^{\lambda i}{\cal Y}_\tau ^{\lambda
i^{^{\prime }}}}}\times   \label{r2.1}
\end{equation}
\[
\left( {\cal Z}^{\lambda ii^{\prime }}\left( \tau \right) \left( -1\right)
^{\lambda +\lambda _1+J^{\prime }}\left( K_1^{J^{\prime }}(\lambda
_1i_2^{\prime },\lambda _1i_1\mid \lambda i,\lambda i^{\prime
})-K_2^{J^{\prime }}(\lambda _1i_2^{\prime },\lambda i\mid \lambda
_1i_1,\lambda i^{\prime })\right) \right. +
\]
\[
\left. \left( \overline{X}_1^{\lambda ii^{\prime }}\left( \tau \right) +%
\overline{X}_1^{\lambda i^{\prime }i}\left( \tau \right) \right)
K^{J^{\prime }}(\lambda _1i_2^{\prime },\lambda i\mid \lambda i^{\prime
},\lambda _1i_1)\right),
\]
}
{\large
\[
\left. {\cal W}_2\right. _{\lambda _1i_1,\lambda
_2i_2}^{\lambda _1^{\prime }i_1^{\prime },\lambda _2^{\prime}i_2^{\prime
}}\left( \lambda \right)=\left( \delta _{i_2^{\prime },i_2}\left. {\cal W}_1
\right. _{i_1}^{i_1^{\prime }}\left( \lambda _1\right)+\delta _{i_1^{\prime },i_1}\left.
{\cal W}_1\right. _{i_2}^{i_2^{\prime }}\left( \lambda _2\right)\right) +
\]
\[
\delta _{\lambda _1^{\prime },\lambda _2}\delta _{\lambda _2^{\prime
},\lambda _1}\left( -1\right) ^{\lambda _1+\lambda _2+\lambda }\left( \delta
_{i_2^{\prime },i_1}\left. {\cal W}_1\right.
_{i_2}^{i_1^{\prime }}\left( \lambda _2\right)+\delta _{i_1^{\prime },i_2}\left. {\cal W}_1
\right. _{i_1}^{i_2^{\prime }}\left(\lambda _1\right)\right) +
\]
\[
\Delta \overline{\omega }^\lambda (\lambda _2^{\prime }i_2^{\prime },\lambda
_1^{\prime }i_1^{\prime }\mid \lambda _1i_1,\lambda _2i_2).
\]
}
The matrix elements (\ref{e2}), (\ref{e3})
have been evaluated by keeping all terms containing the first power of
coefficients $K$, $K_i$  and products $KK$, only.
We took into account also the terms containing the
products $KK_i$ that include the $\psi^3\varphi$ terms.
The other terms including the products $KK_i$ are smaller and they are
neglected.
The terms with products $K_iK_i$ were not taken into account
for these matrix elements because they have no
$\psi^{4}$,$\psi^3\varphi$ terms.
It should be pointed out that the matrix element  (\ref{w1})  differs
from the phonon energy in contrast with the ERPA case
(see Eq.(\ref{ERPA})).
This difference is due to the approximations
(\ref{A1}),(\ref{A2}),(\ref{A3}).
The corrections to the matrix element (\ref{ERPA}) include phonon
amplitudes of the next orders.
Only the terms $h_2$ and  $h_3$ of the hamiltonian (\ref{H1}) contribute
to these corrections.
Then, these corrections become zero if $\varphi=0$
(see the explicit form for $\overline{X}_1$, $K_1$, $K_2$).
It should be mentioned that
we do not use so called quasidiagonal approximation for the
coefficients $K$ and $K_i$.

The values $ \Delta \overline{\omega }^\lambda$ are anharmonic shifts
of energies of the two-phonon configurations due to the Pauli principle
corrections, where
{\large
$
\Delta \overline{\omega }^\lambda =\Delta \overline{\omega }_1^\lambda
+\Delta \overline{\omega }_2^\lambda +\Delta \overline{\omega
}_3^\lambda, $ }

{\large
\begin{equation}
\label{dw1}
\Delta \overline{\omega }_1^\lambda (\lambda _2^{\prime }i_2^{\prime
},\lambda _1^{\prime }i_1^{\prime }\mid \lambda _1i_1,\lambda _2i_2)=
\end{equation}
\[
\left( \omega _{\lambda _1i_1}+\omega _{\lambda _2i_2}\right) K^\lambda
(\lambda _2^{\prime }i_2^{\prime },\lambda _1^{\prime }i_1^{\prime }\mid
\lambda _1i_1,\lambda _2i_2)-
\]
\[
\frac 14\sum_{i\tau }\left( {\large \frac{\overline{X}^{\lambda
_1^{\prime }i_1^{\prime }i}\left( \tau \right) +\overline{X}^{\lambda
_1^{\prime }ii_1^{\prime }}\left( \tau \right) }{\sqrt{{\cal Y}_\tau
^{\lambda _1^{\prime }i_1^{\prime }}{\cal Y}_\tau ^{\lambda _1^{\prime }i}}}}%
K^\lambda (\lambda _2^{\prime }i_2^{\prime },\lambda _1^{\prime }i\mid
\lambda _1i_1,\lambda _2i_2)+\right.
\]
\[
+\frac{\overline{X}^{\lambda _2^{\prime }i_2^{\prime }i}\left( \tau \right) +%
\overline{X}^{\lambda _2^{\prime }ii_2^{\prime }}\left( \tau \right) }{\sqrt{%
{\cal Y}_\tau ^{\lambda _2^{\prime }i_2^{\prime }}{\cal Y}_\tau ^{\lambda
_2^{\prime }i}}}K^\lambda (\lambda _1^{\prime }i_1^{\prime },\lambda
_2^{\prime }i\mid \lambda _1i_1,\lambda _2i_2)\left( -1\right) ^{\lambda
_1^{\prime }+\lambda _2^{\prime }+\lambda }+
\]
\[
\sum_{\lambda _4\lambda _3i_3i^{\prime }}\left( \frac{\overline{X}^{\lambda
_4ii^{\prime }}\left( \tau \right) +\overline{X}^{\lambda _4i^{\prime
}i}\left( \tau \right) }{\sqrt{{\cal Y}_\tau ^{\lambda _4i}{\cal Y}_\tau
^{\lambda _4i^{^{\prime }}}}}\right. \times
\]
\[
\left. \left. K^\lambda (\lambda _3i_3,\lambda _4i^{\prime }\mid \lambda
_1i_1,\lambda _2i_2)K^\lambda (\lambda _2^{\prime }i_2^{\prime },\lambda
_1^{\prime }i_1^{\prime }\mid \lambda _4i,\lambda _3i_3)\right) \right),
\]
}

{\large
\begin{equation}
\Delta \overline{\omega }_2^\lambda (\lambda _2^{\prime }i_2^{\prime
},\lambda _1^{\prime }i_1^{\prime }\mid \lambda _1i_1,\lambda _2i_2)=-\frac
18\sum_{i\tau }\left( \frac{{\cal Z}^{\lambda _2i_2i}\left( \tau \right)+
{\cal Z}^{\lambda _2ii_2}\left( \tau \right) }{\sqrt{{\cal Y}_\tau
^{\lambda _2i_2}{\cal Y}_\tau ^{\lambda _2i}}}\times \right.
\label{dw2}
\end{equation}
\[
K_1^\lambda (\lambda _2^{\prime }i_2^{\prime },\lambda _1i_1\mid \lambda
_2i,\lambda _1^{\prime }i_1^{\prime })-\frac{{\cal Z}^{\lambda_1^{\prime
}i_1^{\prime }i}\left( \tau \right) +{\cal Z}^{\lambda _1^{\prime
}ii_1^{\prime }}\left( \tau \right) }{\sqrt{{\cal Y}_\tau ^{\lambda
_1^{\prime }i_1^{\prime }}{\cal Y}_\tau ^{\lambda _1^{\prime }i}}}\times
\]
\[
K_2^\lambda (\lambda _2^{\prime }i_2^{\prime },\lambda _1^{\prime }i\mid
\lambda _1i_1,\lambda _2i_2)-\frac{{\cal Z}^{\lambda _2^{\prime
}i_2^{\prime }i}\left( \tau \right) +{\cal Z}^{\lambda _2^{\prime
}ii_2^{\prime }}\left( \tau \right) }{\sqrt{{\cal Y}_\tau ^{\lambda
_2^{\prime }i_2^{\prime }}{\cal Y}_\tau ^{\lambda _2^{\prime }i}}}\times
\]
\[
\left( -1\right) ^{\lambda +\lambda _1^{\prime }+\lambda _2^{\prime
}}K_2^\lambda (\lambda _1^{\prime }i_1^{\prime },\lambda _2^{\prime }i\mid
\lambda _1i_1,\lambda _2i_2)-
\]
\[
\sum_{\lambda _4\lambda _3i_3i^{\prime }}\left( \frac{{\cal Z}^{\lambda
_4ii^{\prime }}\left( \tau \right) +{\cal Z}^{\lambda _4i^{\prime
}i}\left( \tau \right) }{\sqrt{{\cal Y}_\tau ^{\lambda _4i}{\cal Y}
_\tau ^{\lambda _4i^{^{\prime }}}}}\left( K^\lambda (\lambda _2^{\prime
}i_2^{\prime },\lambda _1^{\prime }i_1^{\prime }\mid \lambda _4i,\lambda
_3i_3)\right. \right. \times
\]
\[
K_2^\lambda (\lambda _3i_3,\lambda _4i^{\prime }\mid \lambda _1i_1,\lambda
_2i_2)-\sum_{J_1J_2}\left( -1\right) ^{\lambda _1+\lambda _2^{\prime
}+J_1+J_2}\times
\]
\[
\left( \left\{
\begin{array}{ccc}
J_1 & \lambda _3 & \lambda _1 \\
\lambda _4 & J_2 & \lambda _2 \\
\lambda _1^{\prime } & \lambda _2^{\prime } & \lambda
\end{array}
\right\} K^{J_1}(\lambda _3i_3,\lambda _1i_1\mid \lambda _4i,\lambda
_1^{\prime }i_1^{\prime })\right. \times
\]
\[
K_1^{J_2}(\lambda _2^{\prime }i_2^{\prime },\lambda _2i_2\mid \lambda
_4i^{\prime },\lambda _3i_3)+\left( -1\right) ^{\lambda _3+\lambda
_1^{\prime }+J_1+J_2+\lambda }\left\{
\begin{array}{ccc}
J_1 & \lambda _3 & \lambda _1 \\
\lambda _4 & J_2 & \lambda _2 \\
\lambda _2^{\prime } & \lambda _1^{\prime } & \lambda
\end{array}
\right\} \times
\]
\[
\left. \left. K^{J_1}(\lambda _3i_3,\lambda _1i_1\mid \lambda _4i,\lambda
_2^{\prime }i_2^{\prime })K_1^{J_2}(\lambda _3i_3,\lambda _2i_2\mid \lambda
_4i^{\prime },\lambda _1^{\prime }i_1^{\prime })\right) \right) +
\]
\[
\left( -1\right) ^{\lambda _2+\lambda _3+\lambda _4+\lambda }\frac{\left(
2\lambda _3+1\right) {\cal Z}^{\lambda _4ii^{\prime }}\left( \tau \right)
}{\left( 2\lambda _1+1\right) \sqrt{{\cal Y}_\tau ^{\lambda
_4i}{\cal Y}_\tau ^{\lambda _4i^{^{\prime }}}}}\times
\]
\[
\left( K_2^{\lambda _3}(\lambda _1i_3,\lambda _4i\mid \lambda _1i_1,\lambda
_4i^{\prime })K^\lambda (\lambda _2^{\prime }i_2^{\prime },\lambda
_1^{\prime }i_1^{\prime }\mid \lambda _2i_2,\lambda _1i_3)+\left( -1\right)
^{\lambda} \left( \frac{2\lambda _1+1}{2\lambda _2+1}\right) \times
\right.
\]
\[
\left. \left. \left. K_2^{\lambda _3}(\lambda _2i_3,\lambda _4i\mid \lambda
_2i_2,\lambda _4i^{\prime })K^\lambda (\lambda _2^{\prime }i_2^{\prime
},\lambda _1^{\prime }i_1^{\prime }\mid \lambda _1i_1,\lambda _2i_3)\right)
\right) \right) ,
\]
}

{\large
\begin{equation}
\label{dw3}
\Delta \overline{\omega }_3^\lambda (\lambda _2^{\prime }i_2^{\prime
},\lambda _1^{\prime }i_1^{\prime }\mid \lambda _1i_1,\lambda _2i_2)=-\frac
14\sum_{i\tau \lambda _3}\left( \frac{\overline{X}^{\lambda _2i_2i}\left(
\tau \right) +\overline{X}^{\lambda _2ii_2}\left( \tau \right) }{\sqrt{{\cal %
Y}_\tau ^{\lambda _2i_2}{\cal Y}_\tau ^{\lambda _2i}}}\times \right.
\end{equation}
\[
K_4^{\lambda _3}(\lambda _2^{\prime }i_2^{\prime },\lambda _2i\mid \lambda
_1i_1,\lambda _1^{\prime }i_1^{\prime })\left( -1\right) ^{\lambda +\lambda
_3+\lambda _2+\lambda _1^{\prime }+1}\left( 2\lambda _3+1\right) \left\{
\begin{array}{ccc}
\lambda _1 & \lambda _2 & \lambda  \\
\lambda _2^{\prime } & \lambda _1^{\prime } & \lambda _3
\end{array}
\right\} +
\]
\[
\frac 12\sum_{\lambda _4i_4i^{\prime}}\frac{\overline{X}_1^{\lambda_4ii^{\prime
}}\left( \tau \right) +\overline{X}_1^{\lambda _4i^{\prime }i}\left( \tau
\right) }{\sqrt{{\cal Y}_\tau ^{\lambda _4i}{\cal Y}_\tau ^{\lambda
_4i^{^{\prime }}}}}\times
\]
\[
\left( \left( \frac{2\lambda _3+1}{2\lambda _1+1}\right) K^{\lambda
_3}(\lambda _1i_4,\lambda _4i\mid \lambda _4i^{\prime },\lambda
_1i_1)K^\lambda (\lambda _2^{\prime }i_2^{\prime },\lambda _1^{\prime
}i_1^{\prime }\mid \lambda _1i_4,\lambda _2i_2){\large +}\right.
\]
\[
\left. \left. \left( \frac{2\lambda _3+1}{2\lambda _2+1}\right) K^{\lambda
_3}(\lambda _2i_4,\lambda _4i\mid \lambda _4i^{\prime },\lambda
_2i_2)K^\lambda (\lambda _2^{\prime }i_2^{\prime },\lambda _1^{\prime
}i_1^{\prime }\mid \lambda _1i_1,\lambda _2i_4)\right) \right).
\]
}
One can prove that
$\Delta \overline{\omega }_1^\lambda\gg\Delta \overline{\omega
}_2^\lambda $ and
$\Delta \overline{\omega }_1^\lambda\gg\Delta \overline{\omega
}_3^\lambda $ in the case $\psi\gg\varphi$.
Moreover, the shifts
$
\Delta \overline{\omega }_2^\lambda
$
and
$
\Delta \overline{\omega }_3^\lambda
$
become zero if $\varphi=0$.
It is necessary to emphasize that only the terms  $h_2$ and $h_3$
 contribute to the shifts
$
\Delta \overline{\omega }_2^\lambda
$
and
$
\Delta \overline{\omega }_3^\lambda
$.
The shifts $\Delta \overline{\omega }_2^\lambda$ and
$\Delta \overline{\omega }_3^\lambda$   have been neglected
in our previous papers.
The matrix elements coupling one- and  two-phonon configurations
(\ref{e4}) are
{\large
\begin{eqnarray*}
\left. {\cal U}\right. _{\lambda _1i_1,\lambda
_2i_2}^{\lambda i}\left( \lambda \right)&=&\sum_\tau \left. U_1\right. _{\lambda _2i_2}^{\lambda
_1i_1}(\lambda i,\tau )+\left. U_2\right. _{\lambda _2i_2}^{\lambda
_1i_1}(\lambda i,\tau )+  \label{r3} \\
&&\ \left( -1\right) ^{\lambda _1+\lambda _2+\lambda }\left. U_2\right.
_{\lambda _1i_1}^{\lambda _2i_2}(\lambda i,\tau )+\overline{U}_{\lambda
_2i_2}^{\lambda _1i_1}(\lambda i,\tau ),
\end{eqnarray*}
}
where
{\large
\[
\left. U_1\right. _{\lambda _2i_2}^{\lambda _1i_1}(\lambda i,\tau )=%
{\normalsize (-1)^{\lambda _1+\lambda _2+\lambda }\frac 1{\sqrt{2}}\sqrt{%
(2\lambda _1+1)(2\lambda _2+1)}\left. \sum_{j_1j_2j_3}\right. ^\tau
(1-q_{j_2j_3})\times }
\]
\[
\frac{f_{j_1j_2}^\lambda }{\sqrt{{\cal Y}_\tau ^{\lambda i}}}\left\{
\begin{array}{ccc}
\lambda _1 & \lambda _2 & \lambda  \\
j_2 & j_1 & j_3
\end{array}
\right\} \left( {\cal L}_{j_1j_2}^{\lambda i}\left( \tau \right) \psi
_{j_2j_3}^{\lambda _2i_2}\varphi _{j_3j_1}^{\lambda _1i_1}+{\cal L}%
_{j_2j_1}^{\lambda i}\left( \tau \right) \psi _{j_3j_1}^{\lambda _1i_1}\varphi
_{j_2j_3}^{\lambda _2i_2}\right),
\]
}
{\large
\[
{\cal L}_{j_1j_2}^{\lambda i}\left( \tau \right) =\frac 12\left(
L_{j_1j_2}^{\lambda i\left( +\right) }+L_{j_2j_1}^{\lambda i\left( -\right)
}\right),
\]
}

{\large
\[
\left. U_2\right. _{\lambda _2i_2}^{\lambda _1i_1}(\lambda i,\tau )=%
{\normalsize (-1)^{\lambda _1+\lambda _2+\lambda }\frac 1{\sqrt{2}}\sqrt{%
(2\lambda _1+1)(2\lambda _2+1)}\left. \sum_{j_1j_2j_3}\right. ^\tau (1-}q%
{\normalsize _{j_2j_3})\times }
\]
\[
\frac{f_{j_1j_2}^{\lambda _2}}{\sqrt{{\cal Y}_\tau ^{\lambda _2i_2}}}\left\{
\begin{array}{ccc}
\lambda _1 & \lambda _2 & \lambda  \\
j_1 & j_3 & j_2
\end{array}
\right\} \left( {\cal L}_{j_1j_2}^{\lambda _2i_2}\left( \tau \right) \varphi
_{j_2j_3}^{\lambda _1i_1}\varphi _{j_3j_1}^{\lambda i}+{\cal L}%
_{j_2j_1}^{\lambda _2i_2}\left( \tau \right) \psi _{j_3j_1}^{\lambda i}\psi
_{j_2j_3}^{\lambda _1i_1}\right),
\]
}

{\large
\begin{equation}
\label{UK}
\overline{U}_{\lambda _2i_2}^{\lambda _1i_1}(\lambda i,\tau)=\sum_{\lambda
^{\prime }i^{\prime }\lambda _3i_3}\left( \left. U_2\right. _{\lambda
_3i_3}^{\lambda ^{\prime }i^{\prime }}(\lambda i,\tau )K^\lambda (\lambda
^{\prime }i^{\prime }\lambda _3i_3\mid \lambda _1i_1,\lambda _2i_2)+\right.
\]
\[
\left. U_3\right. _{\lambda _3i_3}^{\lambda ^{\prime }i^{\prime }}(\lambda
i,\tau )K_1^\lambda (\lambda _2i_2\lambda ^{\prime }i^{\prime }|\lambda
_3i_3\lambda _1i_1,)+\sum_{J^{\prime }}\left( 2J^{\prime }+1\right) \sqrt{%
\frac{2\lambda ^{\prime }+1}{2\lambda +1}}\times
\end{equation}
\[
\left( \left\{
\begin{array}{ccc}
\lambda _1 & \lambda _2 & \lambda  \\
J^{\prime } & \lambda ^{\prime } & \lambda _3
\end{array}
\right\} \left( K^{J^{\prime }}(\lambda ^{\prime }i^{\prime },\lambda i\mid
\lambda _3i_3,\lambda _2i_2)\left. U_1\right. _{\lambda ^{\prime }i^{\prime
}}^{\lambda _1i_1}(\lambda _3i_3,\tau )\left( -1\right) ^{\lambda +\lambda
^{\prime }+\lambda _2+\lambda _3}\right. +\right.
\]
\[
\left. K_1^{J^{\prime }}(\lambda ^{\prime }i^{\prime }\lambda _2i_2\mid
\lambda _3i_3,\lambda i)\left. U_4\right. _{\lambda ^{\prime }i^{\prime
}}^{\lambda _1i_1}(\lambda _3i_3,\tau )\left( -1\right) ^{\lambda +\lambda
^{\prime }+J^{\prime }}\right) {\large +}
\]
\[
{\large \left\{
\begin{array}{ccc}
\lambda _1 & \lambda _2 & \lambda  \\
\lambda ^{\prime } & J^{\prime } & \lambda _3
\end{array}
\right\} }\left( K^{J^{\prime }}(\lambda ^{\prime }i^{\prime },\lambda i\mid
\lambda _3i_3,\lambda _1i_1)\left. U_1\right. _{\lambda _2i_2}^{\lambda
^{\prime }i^{\prime }}(\lambda _3i_3,\tau )+\right.
\]
\[
\left. \left. K_1^{J^{\prime }}(\lambda ^{\prime }i^{\prime }\lambda
_1i_1\mid \lambda _3i_3,\lambda i)\left. U_4\right. _{\lambda
_2i_2}^{\lambda ^{\prime }i^{\prime }}(\lambda _3i_3,\tau )\left( -1\right)
^{\lambda _1+\lambda _3+J^{\prime }}\right) \right)
\]
}

{\large
\[
\left. U_3\right. _{\lambda _2i_2}^{\lambda _1i_1}(\lambda i,\tau )=%
{\normalsize (-1)^{\lambda _1+\lambda _2+\lambda }\frac 1{\sqrt{2}}\sqrt{%
(2\lambda _1+1)(2\lambda _2+1)}\left. \sum_{j_1j_2j_3}\right. ^\tau (1-}q%
{\normalsize _{j_2j_3})\times }
\]
\[
\frac{f_{j_1j_2}^{\lambda _2}}{\sqrt{{\cal Y}_\tau ^{\lambda _2i_2}}}\left\{
\begin{array}{ccc}
\lambda _1 & \lambda _2 & \lambda  \\
j_1 & j_3 & j_2
\end{array}
\right\} \left( {\cal L}_{j_2j_1}^{\lambda _2i_2}\left( \tau \right) \varphi
_{j_2j_3}^{\lambda _1i_1}\varphi _{j_3j_1}^{\lambda i}+{\cal L}%
_{j_1j_2}^{\lambda _2i_2}\left( \tau \right) \psi _{j_3j_1}^{\lambda i}\psi
_{j_2j_3}^{\lambda _1i_1}\right),
\]
}

{\large
\[
\left. U_4\right. _{\lambda _2i_2}^{\lambda _1i_1}(\lambda i,\tau )=%
{\normalsize (-1)^{\lambda _1+\lambda _2+\lambda }\frac 1{\sqrt{2}}\sqrt{%
(2\lambda _1+1)(2\lambda _2+1)}\left. \sum_{j_1j_2j_3}\right. ^\tau (1-}%
q_{j_2j_3})\times
\]
\[
\frac{f_{j_1j_2}^\lambda }{\sqrt{{\cal Y}_\tau ^{\lambda i}}}\left\{
\begin{array}{ccc}
\lambda _1 & \lambda _2 & \lambda  \\
j_2 & j_1 & j_3
\end{array}
\right\} \left( {\cal L}_{j_2j_1}^{\lambda i}\left( \tau \right) \psi
_{j_2j_3}^{\lambda _2i_2}\varphi _{j_3j_1}^{\lambda _1i_1}+{\cal L}%
_{j_1j_2}^{\lambda i}\left( \tau \right) \psi _{j_3j_1}^{\lambda _1i_1}\varphi
_{j_2j_3}^{\lambda _2i_2}\right).
\]
}
The terms including the coefficients $K_1$ in (\ref{UK})
were neglected in the papers \cite{KVCGS98,SVK99,VKCS00}.

Solving the system of equations (\ref{EQ}) we obtain the energies
($E_\nu$)
of excited states and the coefficients
$R_i(J\nu)$ and  $P_{\lambda _2i_2}^{\lambda_1i_1}(J\nu )$ of the wave
function (\ref{WF}).
These equations are more general than ones derived in
\cite{KVCGS98,SVK99,VKCS00}.
They have the same form as the basic QPM equations \cite{VG,VS83}.
The GSC change phonon energies $\omega_{\lambda i}$,
the anharmonic shifts of two-phonon configurations
and the matrix elements coupling one- and  two-phonon configurations.
In the case when $\psi\gg\varphi$ the system of equations (\ref{EQ})
can be reduced to  the system of equations of the extended QPM
\cite{KVCGS98,SVK99,VKCS00}.
If we put $q_{j}=0 $, $\overline{X}_1=0$  and $K_i=0$ we get the
usual  QPM equations with taking into  account the
Pauli principle corrections \cite{VG,VV,VS83}.
In the case when $q_{j}=0 $, $K=0$ and  $K_i=0$
we have equations describing coupling of one-
and two- RPA phonons without taking into account the Pauli principle \cite{VG}.

\section{Anharmonic shifts of two-phonon configurations}
Now we discuss the correspondence between the QPM equations  presented
above and the diagrams of the Nuclear Field Theory (NFT)
\cite{BM75,B77}.
For the sake of simplicity we compare these  approaches for the
hamiltonian
(\ref{1}) including the average nuclear field and the isoscalar
particle-hole residual forces only. Besides that, we put $q_{j}=0 $
in the system of equations (\ref{EQ}), i.e. conventional RPA phonons
are used as the QPM basis in this case.

The NFT is a formulation of many-body perturbation theory with
vibrational modes summed to all orders in RPA. Its building blocks are
RPA phonons and the single particle degrees of freedom which are
described  in the Hartree-Fock  approximation. The coupling between them
is treated diagramatically in the perturbation theory.
Diagrams illustrating first order coupling between the surface vibrations
and the fermion fields are shown in Fig. 1.
The wavy lines are phonon propagations, while the particles and
the holes are depicted by the arrowed lines. The lowest-order
anharmonic terms of NFT contributing to the energy
of two-phonon states are  represented by fourth-order diagrams shown in
Fig. 2 [3,35-41].
%\cite{BM75,H74,M74,BES76,BBH99,H99,PBBV2000}.
These graphs are called butterfly-type (A,B), trapezoid-type (C,D), and
diamond-type (E,F) diagrams.
For each diagram shown in Fig. 2 there are 5 other diagrams which are
obtained by changing the direction of the phonon lines.

One can rewrite the system of equations (\ref{EQ}) in the space of
two-phonon states. The diagonal approximation for the matrix element
$\left. {\cal W}_1\right. _i^{i^{\prime}}\left( \lambda \right)$
 \\ ($
\left. {\cal W}_1\right. _i^{i^{\prime
}}\left( \lambda \right)=\delta _{i,i^{\prime }}\left. {\cal W}_1
\right. _i^i\left( \lambda \right)
$)
enables one to find the coefficient $R_i(J\nu)$
from the first equation of the system (\ref{EQ}).
Substituting  it into the second equation of this system one can get
the following secular equation to find
energies $E_\nu$:
{\large
\[
\det \left\| \left( \delta _{i_2^{\prime },i_2}\left. {\cal W}%
_1\right. _{i_1}^{i_1^{\prime }}\left( \lambda _1\right)+\delta
_{i_1^{\prime },i_1}\left. {\cal W}_1\right.
_{i_2}^{i_2^{\prime }} \left( \lambda _2\right)\right)\delta _{\lambda _1^{\prime },\lambda
_1}\delta _{\lambda _2^{\prime },\lambda _2}+\right.
\]
\[
\delta _{\lambda _1^{\prime },\lambda _2}\delta _{\lambda _2^{\prime
},\lambda _1}\left( -1\right) ^{\lambda _1+\lambda _2+\lambda }\left( \delta
_{i_2^{\prime },i_1}\left. {\cal W}_1\right.
_{i_2}^{i_1^{\prime }}\left( \lambda _2\right)+\delta _{i_1^{\prime },i_2}\left. {\cal W}
_1\right. _{i_1}^{i_2^{\prime }}\left( \lambda _1\right) \right) +
\]
\[
\left. \Delta \overline{\omega }^\lambda (\lambda _2^{\prime }i_2^{\prime
},\lambda _1^{\prime }i_1^{\prime }\mid \lambda _1i_1,\lambda _2i_2)-\Delta
U^\lambda (\lambda _2^{\prime }i_2^{\prime },\lambda _1^{\prime }i_1^{\prime
}\mid \lambda _1i_1,\lambda _2i_2)-E_\nu \right\| =0,
\]
}
where
{\large
\begin{equation}
\label{dU}
\Delta U^\lambda (\lambda _2^{\prime }i_2^{\prime },\lambda _1^{\prime
}i_1^{\prime }\mid \lambda _1i_1,\lambda
_2i_2)=\sum_i\frac{{\cal{U}}_{\lambda _1i_1,\lambda
_2i_2}^{\lambda i}\left( \lambda \right){\cal U}
_{\lambda _1^{\prime }i_1^{\prime },\lambda _2^{\prime
}i_2^{\prime }}^{\lambda i}\left( \lambda \right)}{\left. {\cal W}_1
\right. _i^i\left( \lambda \right)-E_\nu }.
\end{equation}
}
It is seen that the values $\Delta U$ and
$\Delta\overline{\omega}$ are anharmonic shifts of two-phonon states.
The term $h_{QB}$ does not contribute
to the energy shifts of $\Delta\overline{\omega}$, while
it results in the energy shifts of $\Delta U$.
Furthermore, the term $h_{QB}$ contributes to the scattering vertices
(see  Fig.1 (A), (B)) only.
A proof of this statement is similar to one of \cite{BES76}.
For example we consider terms $\Delta K$ (see (\ref{dw1})) and
$\Delta U_2$ (see (\ref{dU})):
{\large
\[
\Delta K^\lambda (\lambda _4 i_4,
\lambda _3 i_3 \mid \lambda _1i_1,\lambda _2i_2)=
\]
\[
- \frac 14\sum_{\tau }\sum_{i_5}\frac{\overline{X}^{\lambda
_3i_3i_5}\left( \tau \right) +\overline{X}^{\lambda
_3i_5i_3}\left( \tau \right)}{\sqrt{{\cal Y}_\tau
^{\lambda _3i_3}{\cal Y}_\tau ^{\lambda _3i_5}}}
K^\lambda (\lambda _4i_4,\lambda _3i_5\mid\lambda _1i_1,\lambda _2i_2),
\]
}
{\large
\[
\Delta U_2^\lambda (\lambda_3 i_3,\lambda _4
i_4\mid \lambda _2i_2,\lambda _1i_1)=\sum_\tau \sum_{i_5}\frac{\left.
U_2\right. _{\lambda _1i_1}^{\lambda _2i_2}(\lambda i_5,\tau )\left.
U_2\right. _{\lambda _3i_3}^{\lambda _4
i_4}(\lambda i_5,\tau )}{\left. {\cal{W}}_1
\right. _{i_5}^{i_5}\left( \lambda\right) -E_\nu }.
\]
}
The anharmonic shifts of energies
$\Delta K^\lambda (\lambda _4 i_4,
\lambda _3 i_3 \mid \lambda _1i_1,\lambda _2i_2)$ and \\
$\Delta U_2^\lambda (\lambda_3 i_3,\lambda _4
i_4\mid \lambda _2i_2,\lambda _1i_1)$
can be illustrated by the QPM diagrams \cite{VG,VV} shown in Fig. 3
 (A) and (B), respectively.
If the intermediate phonon line ($i_5$) in these shifts
are changed by the two-quasiparticle state lines (i.e. $\psi=1$ and
$\varphi=0$),
then these shifts can be represented by the fourth-order diagrams of
the NFT (Fig. 2).
In this case the term $\Delta K$ corresponds to the butterfly-type
diagrams,
while the term $\Delta U_2$ can be presented as  the diagrams of the
trapezoid-type and the diamond-type.
Using the same method one can prove that the shifts
$\Delta\overline{\omega}$ and $\Delta U$ (see
(\ref{dw1}-\ref{dw3},\ref{dU})) correspond to
diagrams shown in Fig.2 (A), (B) and Fig. 2 (C), (D), (E), (F),
respectively.
The shifts $\Delta\overline{\omega}_2$
and  $\Delta \overline{\omega }_3$ correspond to
the butterfly-type diagrams.

\section{Conclusion}
A consistent treatment of the ground state correlations beyond the RPA
including their influence on the pairing and the phonon-phonon coupling
in nuclei is  presented.
A new general system of nonlinear equations for the
quasiparticle phonon model is derived.
It is demonstrated that this system contains as a particular case all
equations derived for the QPM early.
The new additional Pauli principle corrections
resulting in the anharmonic shifts of energies of the
two-phonon configurations are found.
It is shown that  the anharmonic shifts
due to the Pauli principle corrections
correspond with butterfly-type diagrams of
the nuclear field theory,
while the anharmonic shifts due to the matrix elements
coupling one- and  two-phonon configurations  give rise to the
trapezoid-type and diamond-type diagrams of the NFT.

\section*{Acknowledgments}
We are grateful to Profs. W.Nawrocka, A.I.Vdovin
for fruitful discussions.
This work has been partially supported by the Bulgarian Plenipotentiary
grant for 2000.

\section*{Appendix A}
The coefficients of the hamiltonian  (\ref{H1}) are given by the
following expressions:
{\large
\[
\overline{X}^{\lambda ii^{^{\prime }}}\left( \tau \right) =\sqrt{\frac{{\cal %
Y}_\tau ^{\lambda i}}2}\left( D_0^{\lambda i}\left( \tau \right)
+D_{+}^{\lambda i}\left( \tau \right) z_{+}^{\lambda i^{\prime }}\left( \tau
\right) +D_{-}^{\lambda i}\left( \tau \right) z_{-}^{\lambda i^{\prime
}}\left( \tau \right) \right)
\]
}

{\large
\[
{\cal Z}^{\lambda ii^{^{\prime }}}\left( \tau \right) =\sqrt{\frac{{\cal Y}%
_\tau ^{\lambda i}}2}\left( D_0^{\lambda i}\left( \tau \right) \left(
1-t_0^{\lambda i^{\prime }}\left( \tau \right) \right) +F_0^{\lambda
i}\left( \tau \right) +\left( D_{+}^{\lambda i}\left( \tau \right)
+D_{-}^{\lambda i}\left( \tau \right) \right) \times \right.
\]
\[
\left. \left( z_{-}^{\lambda i^{^{\prime }}}\left( \tau \right)
-z_{+}^{\lambda i^{\prime }}\left( \tau \right) \right) +F_1^{\lambda
i}\left( \tau \right) t_2^{\lambda i^{\prime }}\left( \tau \right)
-F_2^{\lambda i}\left( \tau \right) t_1^{\lambda i^{\prime }}\left( \tau
\right) \right) +
\]
\[
{\large \sqrt{\frac{{\cal Y}_\tau ^{\lambda i^{^{\prime }}}}2}}\left(
D_0^{\lambda i^{^{\prime }}}\left( \tau \right) \left( 1+t_0^{\lambda
i}\left( \tau \right) \right) -F_0^{\lambda i^{\prime }}\left( \tau \right)
+\left( D_{-}^{\lambda i^{^{\prime }}}\left( \tau \right) -D_{+}^{\lambda
i^{^{\prime }}}\left( \tau \right) \right) \times \right.
\]
\[
\left. \left( z_{-}^{\lambda i}\left( \tau \right) +z_{+}^{\lambda i}\left(
\tau \right) \right) +t_1^{\lambda i}\left( \tau \right) F_2^{\lambda
i^{\prime }}\left( \tau \right) -t_2^{\lambda i}\left( \tau \right)
F_1^{\lambda i^{\prime }}\left( \tau \right) \right)
\]
}

{\large
\begin{eqnarray*}
\overline{X}_1^{\lambda ii^{^{\prime }}}\left( \tau \right)  &=&\sqrt{\frac{%
{\cal Y}_\tau ^{\lambda i}}2}\left( D_0^{\lambda i}\left( \tau \right)
\left( 2-t_0^{\lambda i^{\prime }}\left( \tau \right) \right) -F_0^{\lambda
i}\left( \tau \right) +\left( D_{-}^{\lambda i}\left( \tau \right)
-D_{+}^{\lambda i}\left( \tau \right) \right) \times \right.  \\
&&\left. \left( z_{-}^{\lambda i^{^{\prime }}}\left( \tau \right)
-z_{+}^{\lambda i^{\prime }}\left( \tau \right) \right) +t_2^{\lambda
i^{\prime }}\left( \tau \right) F_2^{\lambda i}\left( \tau \right)
-t_1^{\lambda i^{\prime }}\left( \tau \right) F_1^{\lambda i}\left( \tau
\right) \right)
\end{eqnarray*}
}

{\large
\[
L_{jj^{^{\prime }}}^{\lambda i\left( \pm \right) }\left( \tau \right)
=v_{jj^{^{\prime }}}^{\left( -\right) }+\left( z_{-}^{\lambda i}\left( \tau
\right) \pm z_{+}^{\lambda i}\left( \tau \right) \right) \left(
u_{jj^{^{\prime }}}^{\left( -\right) }-u_{jj^{^{\prime }}}^{\left( +\right)
}\right)
\]
}

{\large
\[
{\cal Y}_\tau ^{\lambda i}=\frac{2\left( 2\lambda \,+\,1\right) ^2}{\left(
D_0^{\lambda i}\left( \tau \right) \left( \kappa _0^{\left( \lambda
,ph\right) }+\kappa _1^{\left( \lambda ,ph\right) }\right) +D_0^{\lambda
i}\left( -\tau \right) \left( \kappa _0^{\left( \lambda ,ph\right) }-\kappa
_1^{\left( \lambda ,ph\right) }\right) \right) ^2}
\]
}

{\large
\[
z_n^{\lambda i}\left( \tau \right) =\frac{D_n^{\lambda i}\left( \tau \right)
\left( \kappa _0^{\left( \lambda ,pp\right) }+\kappa _1^{\left( \lambda
,pp\right) }\right) +D_n^{\lambda i}\left( -\tau \right) \left( \kappa
_0^{\left( \lambda ,pp\right) }-\kappa _1^{\left( \lambda ,pp\right)
}\right) }{D_0^{\lambda i}\left( \tau \right) \left( \kappa _0^{\left(
\lambda ,ph\right) }+\kappa _1^{\left( \lambda ,ph\right) }\right)
+D_0^{\lambda i}\left( -\tau \right) \left( \kappa _0^{\left( \lambda
,ph\right) }-\kappa _1^{\left( \lambda ,ph\right) }\right) }
\]
}

{\large
$n=\left\{ +,-\right\} $
}

{\large
\[
t_{n_1}^{\lambda i}\left( \tau \right) =\frac{F_{n_1}^{\lambda i}\left( \tau
\right) \left( \kappa _0^{\left( \lambda ,pp\right) }+\kappa _1^{\left(
\lambda ,pp\right) }\right) +F_{n_1}^{\lambda i}\left( -\tau \right) \left(
\kappa _0^{\left( \lambda ,pp\right) }-\kappa _1^{\left( \lambda ,pp\right)
}\right) }{D_0^{\lambda i}\left( \tau \right) \left( \kappa _0^{\left(
\lambda ,ph\right) }+\kappa _1^{\left( \lambda ,ph\right) }\right)
+D_0^{\lambda i}\left( -\tau \right) \left( \kappa _0^{\left( \lambda
,ph\right) }-\kappa _1^{\left( \lambda ,ph\right) }\right) }
\]
}

{\large
$n_1=\left\{ 0,1,2\right\} $
}

{\large
\[
D_0^{\lambda i}\left( \tau \right) =\left. \sum_{jj^{^{\prime }}}\right.
^\tau f_{jj^{^{\prime }}}^\lambda (1\,-q_{jj^{^{\prime }}})u_{jj^{^{\prime
}}}^{\left( +\right) }\left( \psi _{jj^{^{\prime }}}^{\lambda i}+\,\varphi
_{jj^{^{\prime }}}^{\lambda i}\right)
\]
}

{\large
\[
D_{\pm }^{\lambda i}\left( \tau \right) =\left. \sum_{jj^{^{\prime
}}}\right. ^\tau f_{jj^{^{\prime }}}^\lambda (1\,-q_{jj^{^{\prime
}}})v_{jj^{^{\prime }}}^{\left( \pm \right) }\left( \psi _{jj^{^{\prime
}}}^{\lambda i}\mp \,\varphi _{jj^{^{\prime }}}^{\lambda i}\right)
\]
}

{\large
\[
F_0^{\lambda i}\left( \tau \right) =\left. \sum_{jj^{^{\prime }}}\right.
^\tau f_{jj^{^{\prime }}}^\lambda (1\,-q_{jj^{^{\prime }}})u_{jj^{^{\prime
}}}^{\left( +\right) }\psi _{jj^{^{\prime }}}^{\lambda i}
\]
}

{\large
\[
F_1^{\lambda i}\left( \tau \right) =\left. \sum_{jj^{^{\prime }}}\right.
^\tau f_{jj^{^{\prime }}}^\lambda (1\,-q_{jj^{^{\prime }}})\left(
v_{jj^{^{\prime }}}^{\left( +\right) }-v_{jj^{^{\prime }}}^{\left( -\right)
}\right) \psi _{jj^{^{\prime }}}^{\lambda i}
\]
}

{\large
\[
F_2^{\lambda i}\left( \tau \right) =\left. \sum_{jj^{^{\prime }}}\right.
^\tau f_{jj^{^{\prime }}}^\lambda (1\,-q_{jj^{^{\prime }}})\left(
v_{jj^{^{\prime }}}^{\left( +\right) }-v_{jj^{^{\prime }}}^{\left( -\right)
}\right) \varphi _{jj^{^{\prime }}}^{\lambda i}
\]
}

{\large
$v_{jj^{^{\prime }}}^{(\pm )}\,=\,u_ju_{j^{\prime }}\,\pm \,v_jv_{j^{\prime
}}$ $u_{jj^{^{\prime }}}^{(\pm )}=u_jv_{j^{^{\prime }}}\pm
\,v_ju_{j^{^{\prime }}}$
}

\section*{Appendix B}
The coefficients $K$ have the following form:
{\large
\begin{eqnarray*}
%\begin{eqnarray}
K^{\lambda} (\lambda _2^{\prime }i_2^{\prime },\lambda _1^{\prime
}i_1^{\prime }
&\mid &\lambda _1i_1,\lambda _2i_2)=\sum_{\mu _1\mu _2\mu _{1^{\prime }}\mu
_{2^{\prime }}}C_{\lambda _1\mu _1\lambda _2\mu _2}^{\lambda \mu }C_{\lambda
_1^{\prime }\mu _1^{\prime }\lambda _2^{\prime }\mu _2^{\prime }}^{\lambda
\mu }\times   \label{K0} \\  \nonumber
K(\lambda _2^{\prime }\mu _2^{\prime }i_2^{\prime },\lambda _1^{\prime }\mu
_1^{\prime }i_1^{\prime } &\mid &\lambda _1\mu _1i_1,\lambda_2\mu_2i_2),
%\end{eqnarray}
\end{eqnarray*}
}

{\large
\[
K^J(\lambda _2^{\prime }i_2^{\prime },\lambda ^{\prime }i^{\prime }\mid
\lambda i,\lambda _2i_2)=\sqrt{(2\lambda +1)(2\lambda _2+1)(2\lambda
^{\prime }+1)(2\lambda _2^{\prime }+1)}\times
\]
\[
\sum\limits_{j_1j_2j_3j_4}(1-q_{j_1j_2})(-1){\large ^{j_2+j_4+\lambda
_2+\lambda _2^{\prime }+J}}\left( \left\{
\begin{array}{ccc}
j_1 & j_2 & \lambda _2^{\prime } \\
j_4 & j_3 & \lambda ^{\prime } \\
\lambda  & \lambda _2 & J
\end{array}
\right\} \times \right.
\]
\[
\left( \psi _{j_1j_2}^{\lambda _2^{\prime }i_2^{\prime }}\psi
_{j_3j_2}^{\lambda _2i_2}\psi _{j_3j_4}^{\lambda ^{\prime }i^{\prime }}\psi
_{j_1j_4}^{\lambda i}-\varphi _{j_1j_2}^{\lambda _2^{\prime }i_2^{\prime
}}\varphi _{j_3j_2}^{\lambda _2i_2}\varphi _{j_3j_4}^{\lambda ^{\prime
}i^{\prime }}\varphi _{j_1j_4}^{\lambda i}\right) \,+
\]
\[
(-1)^{j_1+j_2+j_3+j_4+\lambda +\lambda _2+\lambda ^{\prime }+\lambda
_2^{\prime }+J}\left\{
\begin{array}{ccc}
\lambda  & \lambda _2 & J \\
j_2 & j_4 & j_3
\end{array}
\right\} \left\{
\begin{array}{ccc}
\lambda ^{\prime } & \lambda _2^{\prime } & J \\
j_2 & j_4 & j_1
\end{array}
\right\} \times
\]
\[
\left. \left( \varphi _{j_1j_2}^{\lambda _2^{\prime }i_2^{\prime }}\varphi
_{j_3j_2}^{\lambda _2i_2}\psi _{j_1j_4}^{\lambda ^{\prime }i^{\prime }}\psi
_{j_3j_4}^{\lambda i}-\psi _{j_1j_2}^{\lambda _2^{\prime }i_2^{\prime }}\psi
_{j_3j_2}^{\lambda _2i_2}\varphi _{j_1j_4}^{\lambda ^{\prime }i^{\prime
}}\varphi _{j_3j_4}^{\lambda i}\right) \right),
\]
}

{\large
%\begin{eqnarray}
\begin{eqnarray*}
\nonumber
K_1^{\lambda}(\lambda _2^{\prime }i_2^{\prime },\lambda _1^{\prime
}i_1^{\prime }
&\mid &\lambda _1i_1,\lambda _2i_2)=\sum_{\mu _1\mu _2\mu _{1^{\prime }}\mu
_{2^{\prime }}}C_{\lambda _1^{\prime }\mu _1^{\prime }\lambda _1-\mu
_1}^{\lambda \mu }C_{\lambda _2\mu _2\lambda _2^{\prime }\mu _2^{\prime
}}^{\lambda \mu }\times  \\  \label{K1}
\left( -1\right) ^{\lambda _1-\mu _1}K_1(\lambda _2^{\prime }\mu _2^{\prime
}i_2^{\prime },\lambda _1^{\prime }\mu _1^{\prime }i_1^{\prime } &\mid
&\lambda _1\mu _1i_1,\lambda _2\mu _2i_2),
\end{eqnarray*}
%\end{eqnarray}
}
{\large
\[
K_1^J(\lambda _2^{\prime }i_2^{\prime },\lambda ^{\prime }i^{\prime }\mid
\lambda i,\lambda _2i_2)=\sqrt{(2\lambda +1)(2\lambda _2+1)(2\lambda
^{\prime }+1)(2\lambda _2^{\prime }+1)}\times
\]
\[
\sum\limits_{j_1j_2j_3j_4}(1-q_{j_3j_4})(-1)^{j_1+j_3+\lambda +\lambda
_2^{\prime }}\left\{
\begin{array}{ccc}
\lambda _2 & \lambda _2^{\prime } & J \\
j_4 & j_2 & j_3
\end{array}
\right\} \times
\]
\[
\left( \left( -1\right) ^J\left\{
\begin{array}{ccc}
\lambda ^{\prime } & \lambda  & J \\
j_2 & j_4 & j_1
\end{array}
\right\} \ \left( \psi _{j_4j_1}^{\lambda ^{\prime }i^{\prime }}\psi
_{j_2j_1}^{\lambda i}\varphi _{j_2j_3}^{\lambda _2i_2}\psi
_{j_4j_3}^{\lambda _2^{\prime }i_2^{\prime }}-\varphi _{j_4j_1}^{\lambda
^{\prime }i^{\prime }}\varphi _{j_2j_1}^{\lambda i}\psi _{j_2j_3}^{\lambda
_2i_2}\varphi _{j_4j_3}^{\lambda _2^{\prime }i_2^{\prime }}\right) +\right.
\]
\[
\left. \left\{
\begin{array}{ccc}
\lambda ^{\prime } & \lambda  & J \\
j_4 & j_2 & j_1
\end{array}
\right\} \left( \psi _{j_2j_1}^{\lambda ^{\prime }i^{\prime }}\psi
_{j_4j_1}^{\lambda i}\psi _{j_2j_3}^{\lambda _2i_2}\varphi
_{j_4j_3}^{\lambda _2^{\prime }i_2^{\prime }}-\varphi _{j_2j_1}^{\lambda
^{\prime }i^{\prime }}\varphi _{j_4j_1}^{\lambda i}\varphi
_{j_2j_3}^{\lambda _2i_2}\psi _{j_4j_3}^{\lambda _2^{\prime }i_2^{\prime
}}\right) \right),
\]
}

{\large
%\begin{equation}
\[
\label{K2}
K_2^{\lambda}(\lambda _2^{\prime }i_2^{\prime },\lambda _1^{\prime }i_1^{\prime
}\mid \lambda _1i_1,\lambda _2i_2)=\sum_{\mu _1\mu _2\mu _{1^{\prime }}\mu
_{2^{\prime }}}C_{\lambda _1\mu _1\lambda _2\mu _2}^{\lambda \mu }C_{\lambda
_1^{\prime }-\mu _1^{\prime }\lambda _2^{\prime }\mu _2^{\prime }}^{\lambda
\mu }\times
%\end{equation}
\]
\[
\left( -1\right) ^{\lambda _1^{\prime }-\mu _1^{\prime }-1}K_2(\lambda
_2^{\prime }\mu _2^{\prime }i_2^{\prime },\lambda _1^{\prime }\mu _1^{\prime
}i_1^{\prime }\mid \lambda _1\mu _1i_1,\lambda _2\mu _2i_2),
\]
}

{\large
\[
K_2^J(\lambda _2^{\prime }i_2^{\prime },\lambda ^{\prime }i^{\prime }\mid
\lambda i,\lambda _2i_2)=\sqrt{(2\lambda +1)(2\lambda _2+1)(2\lambda
^{\prime }+1)(2\lambda _2^{\prime }+1)}\times
\]
\[
\sum\limits_{j_1j_2j_3j_4}(1-q_{j_1j_2})(-{\large 1)^{j_2+j_4+\lambda
_2+\lambda _2^{\prime }+J}}\left( \left\{
\begin{array}{ccc}
j_1 & j_2 & \lambda _2^{\prime } \\
j_4 & j_3 & \lambda ^{\prime } \\
\lambda  & \lambda _2 & J
\end{array}
\right\} \times \right.
\]
\[
\left( \psi _{j_1j_2}^{\lambda _2^{\prime }i_2^{\prime }}\psi
_{j_3j_2}^{\lambda _2i_2}\varphi _{j_3j_4}^{\lambda ^{\prime }i^{\prime
}}\psi _{j_1j_4}^{\lambda i}-\varphi _{j_1j_2}^{\lambda _2^{\prime
}i_2^{\prime }}\varphi _{j_3j_2}^{\lambda _2i_2}\psi _{j_3j_4}^{\lambda
^{\prime }i^{\prime }}\varphi _{j_1j_4}^{\lambda i}\right) \,{\large +}
\]
\[
\ (-1)^{j_1+j_2+j_3+j_4+\lambda +\lambda _2+\lambda ^{\prime }+\lambda
_2^{\prime }+J}\left\{
\begin{array}{ccc}
\lambda  & \lambda _2 & J \\
j_2 & j_4 & j_3
\end{array}
\right\} \left\{
\begin{array}{ccc}
\lambda ^{\prime } & \lambda _2^{\prime } & J \\
j_2 & j_4 & j_1
\end{array}
\right\} \times
\]
\[
\left. \left( \varphi _{j_1j_2}^{\lambda _2^{\prime }i_2^{\prime }}\varphi
_{j_3j_2}^{\lambda _2i_2}\varphi _{j_1j_4}^{\lambda ^{\prime }i^{\prime
}}\psi _{j_3j_4}^{\lambda i}-\psi _{j_1j_2}^{\lambda _2^{\prime }i_2^{\prime
}}\psi _{j_3j_2}^{\lambda _2i_2}\psi _{j_1j_4}^{\lambda ^{\prime }i^{\prime
}}\varphi _{j_3j_4}^{\lambda i}\right) \right),
\]
}

{\large
%\begin{equation}
\[
\label{K4}
K_4^{\lambda}(\lambda _2^{\prime }i_2^{\prime },\lambda _1^{\prime
}i_1^{\prime
}\mid \lambda _1i_1,\lambda _2i_2)=\sum_{\mu _1\mu _2\mu _{1^{\prime }}\mu
_{2^{\prime }}}C_{\lambda _1\mu _1\lambda _2-\mu _2}^{\lambda \mu
}C_{\lambda _1^{\prime }-\mu _1^{\prime }\lambda _2^{\prime }\mu _2^{\prime
}}^{\lambda \mu }\times
%\end{equation}
\]
\[
\left( -1\right) ^{\lambda _1^{\prime }-\mu _1^{\prime }+\lambda _2-\mu
_2-1}K_4(\lambda _2^{\prime }\mu _2^{\prime }i_2^{\prime },\lambda
_1^{\prime }\mu _1^{\prime }i_1^{\prime }\mid \lambda _1\mu _1i_1,\lambda
_2\mu _2i_2),
\]
}

{\large
\[
K_4^J(\lambda _2^{\prime }i_2^{\prime },\lambda ^{\prime }i^{\prime }\mid
\lambda i,\lambda _2i_2)=\sqrt{(2\lambda +1)(2\lambda _2+1)(2\lambda
^{\prime }+1)(2\lambda _2^{\prime }+1)}\times
\]
\[
\sum\limits_{j_1j_2j_3j_4}(1-q_{j_1j_2})(-1)^{j_2+j_4+\lambda _2+\lambda
_2^{\prime }+J}\left( \left\{
\begin{array}{ccc}
j_1 & j_2 & \lambda _2^{\prime } \\
j_4 & j_3 & \lambda ^{\prime } \\
\lambda  & \lambda _2 & J
\end{array}
\right\} \times \right.
\]
\[
\left( \psi _{j_1j_2}^{\lambda _2^{\prime }i_2^{\prime }}\varphi
_{j_3j_2}^{\lambda _2i_2}\varphi _{j_3j_4}^{\lambda ^{\prime }i^{\prime
}}\psi _{j_1j_4}^{\lambda i}-\varphi _{j_1j_2}^{\lambda _2^{\prime
}i_2^{\prime }}\psi _{j_3j_2}^{\lambda _2i_2}\psi _{j_3j_4}^{\lambda
^{\prime }i^{\prime }}\varphi _{j_1j_4}^{\lambda i}\right) \,+
\]
\[
(-1)^{j_1+j_2+j_3+j_4+\lambda +\lambda _2+\lambda ^{\prime }+\lambda
_2^{\prime }+J}\left\{
\begin{array}{ccc}
\lambda  & \lambda _2 & J \\
j_2 & j_4 & j_3
\end{array}
\right\} \left\{
\begin{array}{ccc}
\lambda ^{\prime } & \lambda _2^{\prime } & J \\
j_2 & j_4 & j_1
\end{array}
\right\} \times
\]
\[
\left. \left( \varphi _{j_1j_2}^{\lambda _2^{\prime }i_2^{\prime }}\psi
_{j_3j_2}^{\lambda _2i_2}\varphi _{j_1j_4}^{\lambda ^{\prime }i^{\prime
}}\psi _{j_3j_4}^{\lambda i}-\psi _{j_1j_2}^{\lambda _2^{\prime }i_2^{\prime
}}\varphi _{j_3j_2}^{\lambda _2i_2}\psi _{j_1j_4}^{\lambda ^{\prime
}i^{\prime }}\varphi _{j_3j_4}^{\lambda i}\right) \right).
\]
}

%\newpage

\Large
%\newpage
\begin{figure}
\epsfxsize=15cm
\epsfysize=15cm
\centerline{\epsffile{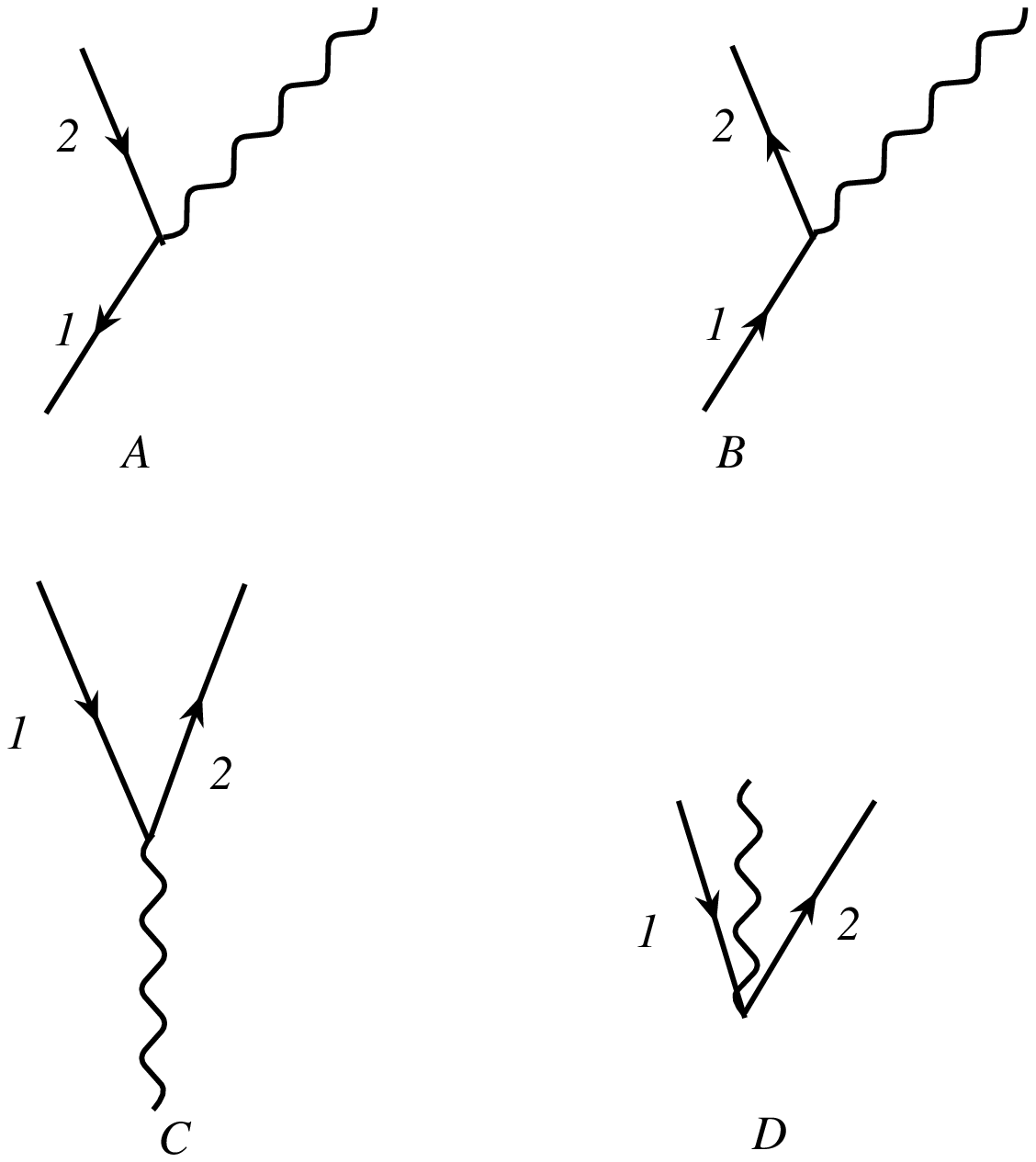}}
\vspace*{6 cm}
\caption{Diagrams illustrating the first order coupling between the
surface vibrations (wavy lines) and the fermion fields (arrowed lines).}
\end{figure}
\vspace*{- 3 cm}
\begin{figure}
\epsfxsize=24cm
\epsfysize=24cm
\vspace*{-2 cm}
\centerline{\epsffile{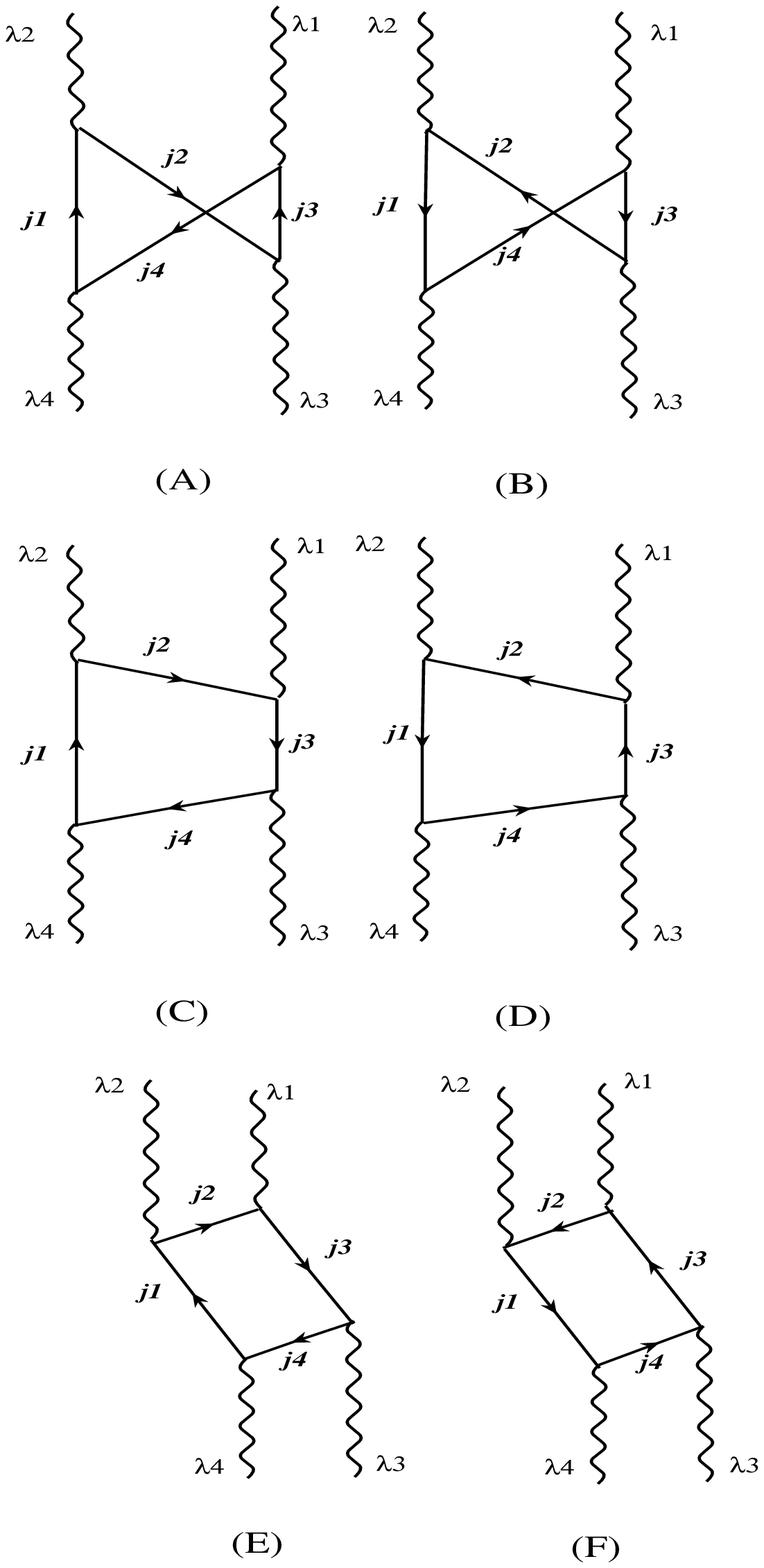}}
\vspace*{-2 cm}
\caption{The NFT diagrams describing  the lowest-order interaction
between two phonons}
\end{figure}
\vspace*{- 3 cm}
\begin{figure}
\epsfxsize=15cm
\epsfysize=17cm
\vspace*{-2 cm}
\centerline{\epsffile{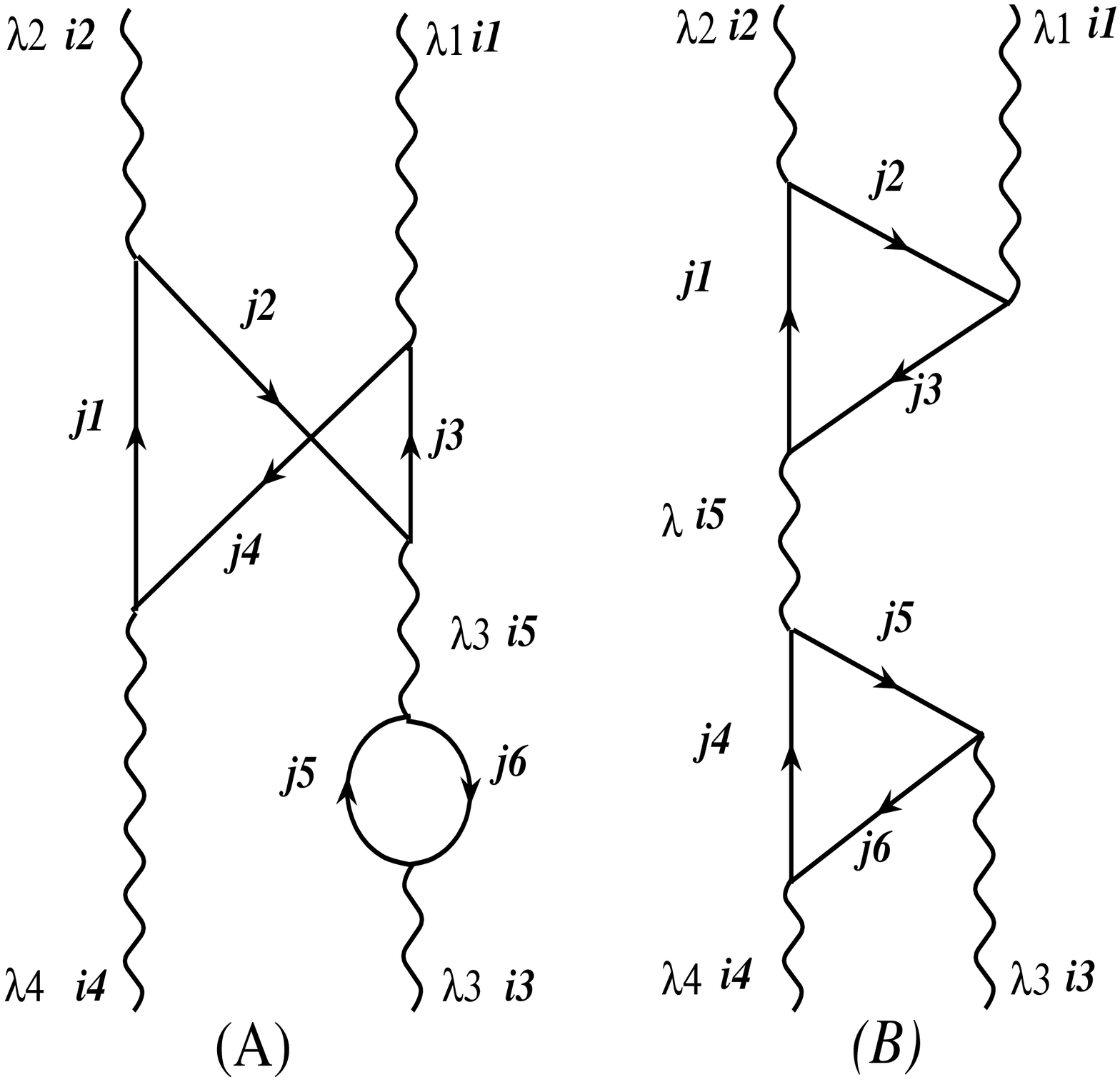}}
\vspace*{-2 cm}
\caption{The QPM diagrams (A) and (B)  illustrating the anharmonic
shifts
$\Delta K^\lambda (\lambda _4 i_4,
\lambda _3 i_3 \mid \lambda _1i_1,\lambda _2i_2)$ and
$\Delta U_2^\lambda (\lambda_3 i_3,\lambda _4
i_4\mid \lambda _2i_2,\lambda _1i_1)$
of energies, respectively.}
\end{figure}
\end{document}